\documentclass[aps,prl,showpacs,twocolumn,epsfig]{revtex4}
\usepackage[dvips]{graphicx}
\begin{document}

\draft
\title{Difference between 
interaction cross sections and reaction cross sections} 

\author{Akihisa Kohama,$^1$ 
Kei Iida,$^{1,2}$ and Kazuhiro Oyamatsu$^{1,3}$}
\affiliation{$^1$RIKEN Nishina Center, RIKEN, 
2-1 Hirosawa, Wako-shi, Saitama 351-0198, Japan\\
$^2$Department of Natural Science, Kochi University, 
Kochi 780-8520, Japan \\
$^3$Department of Media Theories and Production, Aichi Shukutoku
University, Nagakute, Nagakute-cho, Aichi-gun, Aichi 480-1197, Japan
}

\date{\today}

\begin{abstract}
We revisit the commonly accepted notion that the difference 
between interaction and reaction cross sections is negligible at 
relativistic energies, and show that, 
especially in small mass number region, 
it is large enough to help probe nuclear structure. 
For analyses of the difference, we construct ``pseudo data"
for the reaction cross sections using a phenomenological
black-sphere model of nuclei since empirical data are
very limited at high energies.
The comparison with the empirical interaction cross sections 
suggests a significant difference between the reaction and 
interaction cross sections for {\em stable} 
projectiles on a carbon target, which is 
of the order of $0-100$ mb.
\end{abstract}

\pacs{25.60.Dz, 21.10.Gv, 24.10.Ht, 25.40.Cm}  

\maketitle


Measurements of interaction cross sections have been 
done for stable and light unstable nuclei \cite{Ozawa} 
and are planned for heavy unstable 
nuclei in radioactive ion beam facilities, 
such as RIKEN RI Beam Factory.
The interaction cross section, $\sigma_{\rm I}$, 
for a nucleus incident on a target nucleus 
is defined as the total cross section 
for all processes associated with proton and/or 
neutron removal from the incident nucleus \cite{Taniha85}, 
which is measured by a transmission-type experiment. 
In this experiment, the cross section is obtained as
\begin{equation}
   \sigma_{\rm I} = (1/N_t) \log(\gamma_0/\gamma), 
   \label{sigint}
\end{equation}
where $\gamma$ is the ratio of the number of non-interacting
nuclei to the number of incoming nuclei 
for a target-in run, $\gamma_0$ is the same ratio 
for an empty-target run, and $N_t$ is the number of 
the target nuclei per cm$^2$ \cite{Taniha85}.

The above definition of $\sigma_{\rm I}$ leads to the relation, 
$\sigma_{\rm I} = \sigma_{\rm R} - \sigma_{\rm inel}$, 
where $\sigma_{\rm R}$ is the total reaction cross section, and 
$\sigma_{\rm inel}$ is the cross section for 
inelastic channels as will be specified below. 
The total reaction cross section in turn satisfies the relation 
$\sigma_{\rm R} = \sigma_{\rm T} - \sigma_{\rm el}$, 
where $\sigma_{\rm T}$ is the total cross section, and 
$\sigma_{\rm el}$ is the total elastic cross section.

In the measurements of $\sigma_{\rm I}$, 
only the number of events in which an incoming nucleus
has at least one nucleon removed is counted. 
The following processes are {\it not} counted in measuring
$\sigma_{\rm I}$: 
1) Incident nuclei are excited without changing the original
$Z$ and $N$, and target nuclei can change in any way.
2) Target nuclei are excited without changing the
original $Z$ and $N$, while incoming nuclei  
remain in the ground state.
3) Target nuclei break up, while incoming nuclei 
remain in the ground state.
These processes contribute exclusively to
$\sigma_{\rm inel}$ and thus are included in $\sigma_{\rm R}$.
For incident nuclei with no excited bound states, 
such as $^{11}$Li \cite{Taniha92}, 
we expect $\sigma_{\rm inel} \cong 0$ and thus
$\sigma_{\rm I} \cong \sigma_{\rm R}$.

In this Letter, we address the question of how
large the difference between reaction and interaction 
cross sections is.  We systematically analyze empirical 
data for the reaction and interaction cross sections 
measured at high beam energy, $\gtrsim800$ MeV, per 
nucleon.  We complement the limited data for $\sigma_{\rm R}$ 
by constructing pseudo data with the help of a black-sphere 
picture of nuclei \cite{BS1,BS2}.

Theoretically, Ogawa {\it et al}.\ pointed out that, 
for reactions of $^{11}$Li with several kinds of targets,
the contribution of $\sigma_{\rm inel}$ to $\sigma_{\rm R}$ 
is negligibly small \cite{ogawa}.   
Recently, Ozawa {\it et al.} 
experimentally estimated $\sigma_{\rm inel}$ 
for $^{34}$Cl incident on a C target 
as less than about 10 mb \cite{Ozawa02}. 
Since $\sigma_{\rm I} = 1334 \pm 28$ mb, the contribution 
of $\sigma_{\rm inel}$ to $\sigma_{\rm R}$ 
is also negligibly small. 
In both cases, however, the projectiles are loosely-bound systems.
For stable nuclei, whether the contribution 
of $\sigma_{\rm inel}$ to $\sigma_{\rm R}$ is negligibly 
small or not is still an open question.


     Recently, for the purpose of deducing nuclear
size from proton-nucleus elastic scattering and reaction
cross sections, 
we proposed a model in which a nucleus is viewed 
as a ``black'' sphere of radius $a$ \cite{BS1,BS2}.
Here we assume that the target nucleus is strongly 
absorptive to the incident proton and hence acts like 
a black sphere.  
Another requirement for the black-sphere picture is that 
the proton wave length is considerably smaller than the nuclear size.
For proton incident energies higher than about 800 MeV,
these requirements are basically satisfied.

     In this scheme, we first evaluate the black-sphere radius, $a$,
from the measured elastic diffraction peak and then identify $a$ as 
a typical length scale characterizing the nuclear 
size \cite{BS1}.  The center-of-mass (c.m.) scattering angle 
for proton elastic scattering is generally given by 
$\theta_{\rm c.m.} = 2\sin^{-1}(q/2p)$ with the momentum 
transfer, ${\bf q}$, and the proton incident momentum 
in the c.m.\ frame, ${\bf p}$.  For the proton diffraction 
by a circular black disk of radius $a$, we can calculate the 
value of $\theta_{\rm c.m.}$ 
at the first peak as a function of $a$.  
(Here we define the zeroth peak as that whose angle corresponds 
to $\theta_{\rm c.m.}=0$.)  We determine $a$ in such a way that 
this value of $\theta_{\rm c.m.}$ agrees with the first peak angle 
for the measured diffraction in proton-nucleus elastic scattering, 
$\theta_M$.  The radius, $a$, and the angle, $\theta_M$, are then 
related by
\begin{equation}
   2 p a \sin(\theta_M/2) = 5.1356 \cdots.
    \label{a}
\end{equation}

For scattering of protons having energies higher than 
$\sim800$ MeV with stable nuclei, we obtained the following 
results \cite{BS1,BS2}:
1) the absorption cross section, $\pi a^2$, agrees 
with the empirical total reaction cross section within error bars. 
Therefore,
$a$ can be regarded as a ``reaction radius,'' inside 
which the reaction with incident protons occurs.
2) $\sqrt{3/5}a$ ($= r_{\rm BS}$) almost completely 
agrees with the empirically deduced values 
of the root-mean-square matter radius 
for nuclei having mass $A\gtrsim50$,   
while it systematically deviates from the deduced values 
for $A\lesssim50$. 
We also found that, for stable nuclei ranging from He to Pb, 
the black-sphere radius scales as \cite{BS2}
\begin{equation}
  a \simeq 1.2135 A^{1/3} ~[{\rm fm}].
\label{ascale}
\end{equation}

     From the scale $a$ determined above, we calculate 
nucleus-nucleus absorption cross sections, 
which are to be compared 
with empirical total reaction cross sections, $\sigma_{\rm R}$. 
We simply set
\begin{equation}
   \sigma_{\rm BS} = \pi(a_{\rm P} + a_{\rm T})^2, 
   \label{sigbs}
\end{equation}
where $a_{\rm P} (a_{\rm T})$ 
is the black-sphere radius of a projectile (target) nucleus. 
Here we assume that the incident protons 
are point particles as in Ref.~\cite{BS2}.
By substituting the values of $a_{\rm P}$ and $a_{\rm T}$ 
determined by Eq.\ (\ref{a}) into Eq.\ (\ref{sigbs}), 
we evaluate $\sigma_{\rm BS}$ 
for various sets of stable nuclei.  
Expression~(\ref{sigbs}) is merely an assumption, but 
several available data support its validity as we will show later.


Now we concentrate on the reactions of stable projectile nuclei 
on a carbon target.  
Then, Eq.~(\ref{sigbs}) reduces to
\begin{equation}
   \sigma_{\rm BS} = \pi \left(a_{\rm P} + a({\rm C}) \right)^2, 
\label{sigbs2}
\end{equation}
where $a({\rm C})$ is the black-sphere radius of 
the target C nucleus obtained from the measured angle of 
the first diffraction maximum in proton elastic scattering 
\cite{BS2}.  
For proton incident energy higher than $\sim800$ MeV,
$a({\rm C}) = 2.69 \pm 0.07$ fm.  
For later convenience, we introduce 
the interaction radius, $a_{\rm I}$, 
through the following expression:
\begin{equation}
   \sigma_{\rm I} = \pi \left(a_{\rm I} + a({\rm C}) \right)^2.
\label{sigi}
\end{equation}

\begin{figure}[t]
\begin{center}
\includegraphics[width=7cm]{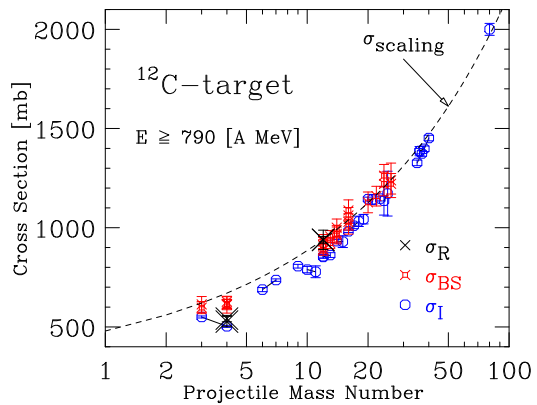}
\end{center}
\vspace{-0.5cm}
\caption{(Color online) 
Comparison of the total reaction cross sections, 
$\sigma_{\rm R}$ ($\times$), 
and their substitutes $\sigma_{\rm BS}$ (squares) 
and $\sigma_{\rm scaling}$ (dashed curve) 
with the interaction cross sections, $\sigma_{\rm I}$ ($\circ$), 
for stable projectiles of $A\le80$ and a $^{12}$C target.  
The absorption cross section, $\sigma_{\rm BS}$, 
defined by Eq.~(\ref{sigbs2}) and the scaling cross section, 
$\sigma_{\rm scaling}$, defined by Eq.~(\ref{sigbss}) 
act as the pseudo data for $\sigma_{\rm R}$.
The empirical data for $\sigma_{\rm R}$ 
are taken from Ref.~\cite{Jaros} (see text), 
and those for
$\sigma_{\rm I}$ are taken from Refs.~\cite{Ozawa,Ozawa02,yamagu}
for a projectile of $^{3, 4}$He, $^{6, 7}$Li, $^9$Be, $^{10, 11}$B, 
$^{12, 13}$C, $^{14, 15}$N, $^{16, 17, 18}$O, 
$^{19}$F, $^{20, 21}$Ne, $^{23}$Na, $^{24, 25}$Mg,
$^{35, 37}$Cl, $^{36, 38, 40}$Ar, and $^{80}$Kr 
at energy per nucleon $\gtrsim800$ MeV.
This figure is the updated version of Fig.~4 of Ref.~\cite{BS2}. 
}
\label{sigrvsi}
\end{figure}

In Fig.~\ref{sigrvsi}, we plot the empirical $\sigma_{\rm R}$ and  
$\sigma_{\rm I}$ data for incident energy per nucleon above 
$\sim800$ MeV.  For comparison, we also plot $\sigma_{\rm BS}$. 
Since the number of the $\sigma_{\rm R}$ data 
is very limited in the energy region of interest here \cite{exfor}, 
we consider $\sigma_{\rm BS}$ as 
pseudo data for $\sigma_{\rm R}$. 
$\sigma_{\rm BS}$ is useful for predicting $\sigma_{\rm R}$
for nuclides for which proton elastic scattering data 
are available while no data are available for $\sigma_{\rm R}$.  
The dashed curve in the figure shows the scaling cross section, 
$\sigma_{\rm scaling}$, for a nucleus-$^{12}$C reaction, 
defined on the basis of Eq.\ (\ref{ascale}) as
\begin{eqnarray}
   \sigma_{\rm scaling} 
   = \pi \left(1.2135 A^{1/3} + a({\rm C}) \right)^2 ~[{\rm fm}^2], 
   \label{sigbss}
\end{eqnarray}
where $a({\rm C})$ is fixed at $2.6930$ fm.
When the data for proton elastic scattering are not available, 
we adopt this $\sigma_{\rm scaling}$ as the pseudo data.

In fact, for incident energies per nucleon higher 
than $\sim800$ MeV, only a few data are available for nucleus-carbon 
total reaction cross sections.  
For the $\sigma_{\rm R}$ data for $^{12}$C$+^{12}$C 
at 870 MeV per nucleon, one finds $939 \pm 17$ mb and 
$939 \pm 49$ mb from Ref.\ \cite{Jaros}. 
By substituting these values into $\sigma_{\rm BS}$ 
in Eq.\ (\ref{sigbs}), one obtains 
$a_{\rm P} = a_{\rm T} = 2.73 \pm 0.03$ fm 
and $2.73 \pm 0.07$ fm, respectively.  
Note that this result is 
consistent with the value of $a({\rm C})$ determined from proton 
elastic scattering data.
For the $\sigma_{\rm I}$ data for the same reacting system, 
on the other hand, 
one finds $856 \pm 9$ mb at 790 MeV per nucleon 
and $853 \pm6$ mb at 950 MeV per nucleon from Ref.\ \cite{Ozawa} .
From Eq.\ (\ref{sigi}) one then obtains 
$a_{\rm I} = 2.53 \pm 0.10$ fm and $2.52 \pm 0.09$ fm. 
The difference between $a({\rm C})$ and this $a_{\rm I}$ 
is about 0 -- 0.3 fm, 
which is typically of the order 
of neutron skin thickness for stable nuclei \cite{Bat:ANP}.
When we discuss the nuclear surface structure, therefore,
such difference should be considered seriously. 

As for the case of $^4$He+$^{12}$C, 
both $\sigma_{\rm BS}$ and $\sigma_{\rm scaling}$
significantly overestimate the empirical values of 
$\sigma_{\rm R}$ ($542 \pm 16$ mb and $527 \pm 26$ mb) \cite{Jaros}
and hence are not acceptable as the pseudo data 
for $\sigma_{\rm R}$.
This exceptional behavior is attributable to the fact that 
excitations associated with internucleon motion are highly 
suppressed in $\alpha$ particles \cite{IKO}.

One can see from Fig.~\ref{sigrvsi} that $\sigma_{\rm I}$ 
is close to $\sigma_{\rm BS}$ in magnitude for the whole
range of the projectile mass, but some deviations do exist. 
In Ref.~\cite{BS2}, we simply stressed a good agreement 
of $\sigma_{\rm BS}$ with $\sigma_{\rm I}$, 
while here we focus on the difference between these two.

\begin{figure}[t]
\begin{center}
\includegraphics[width=7cm]{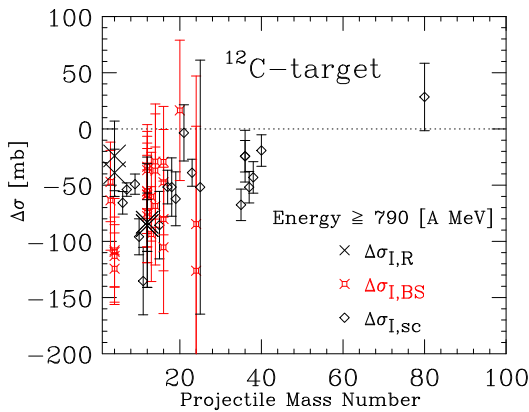}
\end{center}
\vspace{-0.5cm}
\caption{(Color online) 
$\Delta \sigma_{\rm I,R}$ 
$= \sigma_{\rm I} - \sigma_{\rm R}$ ($\times$), 
$\Delta \sigma_{\rm I,BS}$ 
$= \sigma_{\rm I} - \sigma_{\rm BS}$ (squares), and 
$\Delta \sigma_{\rm I,sc}$ 
$= \sigma_{\rm I} - \sigma_{\rm scaling}$ (diamonds) 
as a function of projectile mass.
}
\label{diff}
\end{figure}

In order to clarify the difference, 
we plot $\sigma_{\rm I} - \sigma_{\rm BS}$ or 
$\sigma_{\rm I} - \sigma_{\rm scaling}$,
according to whether proton scattering data are 
available or not,
for stable projectiles of $A\le80$ and a $^{12}$C target 
in Fig.~\ref{diff}.
From the figure, as expected, we find that the difference is 
mostly negative.  The average of the difference over various
projectiles is about $-60.4$ mb.  
This is the major finding of this Letter. 
Interestingly, the difference seems to decrease 
in magnitude with the projectile mass $A$ 
although the plotted data are rather
sparse and often accompanied by large error bars.
We claim that the difference due to 
the choice of $\sigma_{\rm BS}$ or $\sigma_{\rm scaling}$ 
instead of $\sigma_{\rm R}$ is not any artifact. 
The validity of this choice will be discussed later. 

Let us proceed to analyze the $A$-dependence 
of $\sigma_{\rm I} - \sigma_{\rm BS}$ under
the assumption that $\sigma_{\rm R} = \sigma_{\rm BS}$.
Using Eqs.~(\ref{sigbs2}) and (\ref{sigi}), 
we can express the difference as 
\begin{eqnarray}
   \sigma_{\rm I} - \sigma_{\rm BS} 
   &\simeq& -2 \pi \left(a_{\rm P} + a({\rm C}) \right) \Delta a, 
\label{eqdiff}
\end{eqnarray}
where $\Delta a \equiv a_{\rm P} - a_{\rm I}$.  
The above assumption ensures $\Delta a > 0$.
If $\Delta a$ were independent of $A$, 
$|\sigma_{\rm I} - \sigma_{\rm BS}|$ would grow like $A^{1/3}$ as 
$a_{\rm P}$ behaves as in Eq.~(\ref{ascale}). 
As Fig.~\ref{diff} suggests, however, 
$|\sigma_{\rm I} - \sigma_{\rm R}|$ 
approaches a vanishingly small value 
or at least does not increase with $A$. 
This implies that $\Delta a$ decreases 
no slower than $A^{-1/3}$ as $A$ increases. 
As we shall see, this $A$-dependence has 
implications for nuclear structure, a feature
at odds with the commonly accepted notion
that the difference between interaction 
and reaction cross sections is negligible 
at relativistic energies.

Then, what is the physical implication of the difference, 
$\Delta a$?  
From the aforementioned interpretation of the 
black-sphere radius as a reaction radius for incident protons
and the fact that the 
$np$ total cross section is similar to the $pp$ total cross section
at high incident energy above $\sim800$ MeV, we may assume that 
$a_{\rm P}$ corresponds to a critical radius of a projectile
nucleus inside which reactions occur with nucleons in a target 
C nucleus. 
Note that $a_{\rm P}$ is located in the surface region.  
At a radius of $a_{\rm I}$, which is only slightly smaller than 
$a_{\rm P}$, transfer of incident energy 
into excitations of nucleons inside the projectile nucleus 
has to be more effective than at a radius of 
$a_{\rm P}$ because of more frequent reactions and, eventually,  
enough to induce nucleon emission.  
We thus expect that 
$\Delta a$ has relevance to the energy scale characterizing 
breakup of the projectile nucleus, such as single-particle 
level spacing and separation energies.  In fact, 
the $A$-dependence of $\Delta a$ mentioned above could be a key to 
clarifying what energy scale controls nucleon emission.

We remark that the present discussion is 
not always applicable 
when projectiles are deformed nuclei \cite{yabana}.
In this case, $\sigma_{\rm I}$ could be appreciably 
smaller than $\sigma_{\rm R}$ even if $A$ is relatively large. 
This is because a significant part of $\sigma_{\rm inel}$
($= \sigma_{\rm R} - \sigma_{\rm I}$) comes from the low-lying
rotational excitations of the projectile nucleus.  
Candidates for heavy stable nuclei that are deformed in the ground 
state are $^{80}$Kr, $^{154}$Sm, $^{176}$Yb, etc., but
at least for $^{80}$Kr, the effect is not seen 
as long as we assume $\sigma_{\rm R} = \sigma_{\rm scaling}$.


One may wonder if our arguments based on the assumption, 
$\sigma_{\rm R}$ $\simeq \sigma_{\rm BS}$, and Eq.~(\ref{sigbs})
are valid because of a severe shortage of the
real $\sigma_{\rm R}$ data at incident energy 
above $\sim800$ MeV per nucleon.  Even for the available
data for $^{12}$C+$^{12}$C at 870 MeV per nucleon \cite{Jaros},
which we have to rely heavily on, its validity 
remains to be checked.

\begin{figure}[t]
\begin{center}
\includegraphics[width=7cm]{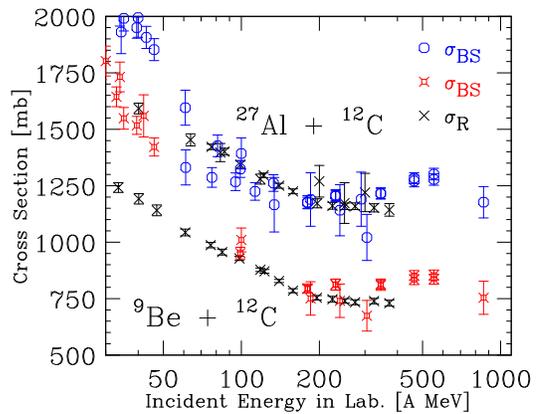}
\end{center}
\vspace{-0.5cm}
\caption{(Color online) 
Comparison of $\sigma_{\rm BS}$ 
(squares for $^{9}$Be+$^{12}$C and
circles for $^{27}$Al+$^{12}$C)
with $\sigma_{\rm R}$ ($\times$)
for the reactions of $^{9}$Be+$^{12}$C (lower) and
$^{27}$Al+$^{12}$C (upper) as a function 
of incident energy per nucleon.
The data of $\sigma_{\rm R}$ are taken 
from Refs.~\cite{exfor,Kox,Takechi}. 
}
\label{bealonc}
\end{figure}

In order to lessen such a concern, 
we proceed to show that $\sigma_{\rm BS}$ given by
Eq.\ (\ref{sigbs2}) does work as
a substitute for $\sigma_{\rm R}$.
On the basis of the fact that $\sigma_{\rm R}$
for proton-nucleus reactions agrees with
$\pi a^2$ within error bars \cite{BS2} and that this tendency
persists for proton incident energy down 
to about 100 MeV \cite{KIO},
we rederive $a_{\rm P}$ and $a({\rm C})$ from 
the corresponding $\sigma_{\rm R}$ ~\cite{Bauho,Auce}
as $(\sigma_{\rm R}/\pi)^{1/2}$,
rather than from proton elastic scattering data. 
(Note that one cannot determine $a_{\rm P}$ 
for C and lighter projectile nuclei and $a({\rm C})$ 
from elastic scattering angular 
distributions measured for the proton incident energies 
less than $\sim400$ MeV, which lack the peak structure.) 
When calculating $a_{\rm P}$ and $a({\rm C})$ to obtain 
$\sigma_{\rm BS}$ for projectile-carbon reactions 
at a given incident energy per nucleon, $T_{\rm P}$, 
we adopt the values of $\sigma_{\rm R}$ 
measured at proton incident energies 
within $\sim 5$ MeV of $T_{\rm P}$.
The obtained values of $\sigma_{\rm BS}$ are to be 
compared with the $\sigma_{\rm R}$ data 
taken with a $^{12}$C target, 
which are, at high incident energy, presently limited 
to such projectiles as $^{9}$Be, $^{12}$C, 
and $^{27}$Al \cite{exfor,Kox,Takechi}.

In Fig.~\ref{bealonc}, we plot $\sigma_{\rm BS}$ and 
$\sigma_{\rm R}$ for the reactions of $^{9}$Be+$^{12}$C and 
$^{27}$Al+$^{12}$C.
The agreement between $\sigma_{\rm BS}$ and $\sigma_{\rm R}$ is 
fairly good for incident energies per nucleon 
ranging $\sim 100-400$ MeV.
Although the uncertainties in $\sigma_{\rm BS}$ are still large,
due mainly to the uncertainties in the measured values of
proton-nucleus total reaction cross sections, 
such agreement strongly supports the effectiveness of 
Eq.\ (\ref{sigbs2}) at predicting $\sigma_{\rm R}$  
for energies per nucleon higher than $\sim 100$ MeV. 
Apparently, the corresponding ratio of $\sigma_{\rm BS}$ to 
$\sigma_{\rm R}$ fluctuates within $\sim10\%$ of unity, but 
if we restrict ourselves to the recent data \cite{Takechi,Auce}, 
the fluctuation becomes much smaller.

Another example is $\sigma_{\rm R}$ versus $\sigma_{\rm BS}$
for $^{12}$C+$^{12}$C reactions. 
In the black-sphere approximation, 
$\sigma_{\rm BS}$ for $^{12}$C+$^{12}$C reactions 
is expressed as $\sigma_{\rm BS} = 4 \pi a({\rm C})^2$, 
because in this case $a_{\rm P} = a(\rm C)$ in Eq.\ (\ref{sigbs2}). 
Let us assume 
that $\pi a({\rm C})^2$ is equal to $\sigma_{\rm R}(p+{\rm C})$, 
where $\sigma_{\rm R}(p+{\rm C})$ is the empirical total
reaction cross section for protons incident 
on a carbon target \cite{Bauho,Auce}.
Then, if $\sigma_{\rm R}$ for $^{12}$C+$^{12}$C reactions
is equal to $4\sigma_{\rm R}(p+{\rm C})$,
we can show that Eq.\ (\ref{sigbs2}) works
also for the case of $^{12}$C+$^{12}$C reactions.

\begin{figure}[t]
\begin{center}
\includegraphics[width=7cm]{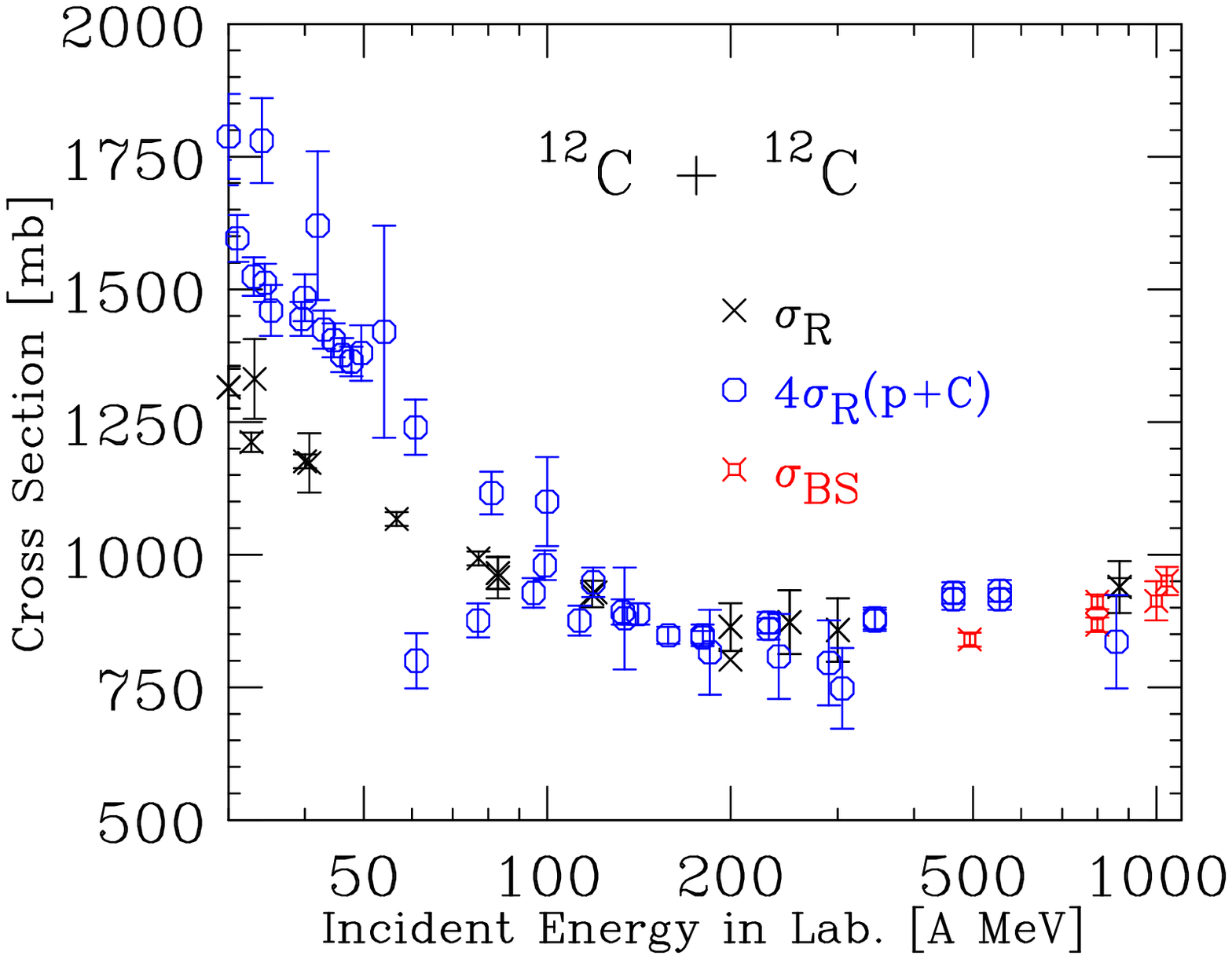}
\end{center}
\vspace{-0.5cm}
\caption{(Color online) 
Comparison of $4\sigma_{\rm R}(p+{\rm C})$ ($\circ$)
with $\sigma_{\rm R}$ for $^{12}$C+$^{12}$C reactions ($\times$)
as a function of incident energy per nucleon.  
The $\sigma_{\rm R}(p+{\rm C})$ data 
are taken from Refs.~\cite{Bauho,Auce}, 
while the $\sigma_{\rm R}$ data for $^{12}$C+$^{12}$C 
reactions are taken from Refs.~\cite{Jaros,exfor,Kox,
Takechi,Kox_npa,Hostachy,Fang00,Zheng,Zhang,Perrin}.  
For reference, we also plot $\sigma_{\rm BS}=4\pi a({\rm C})^2$, 
where $a({\rm C})$ is determined from the measured angle 
of the first 
diffraction peak of proton elastic scattering, by squares.
} 
\label{ccsigr}
\end{figure}

In Fig.~\ref{ccsigr}, we compare $4 \sigma_{\rm R}(p+{\rm C})$
with $\sigma_{\rm R}$ for $^{12}$C+$^{12}$C reactions 
as a function of incident energy per nucleon.
For incident energies per nucleon higher than $\sim100$ MeV,
we obtain an excellent agreement between them. 
This implies the validity of the black-sphere picture
based on the empirical relation 
$\sigma_{\rm R} \cong \sigma_{\rm BS}$ for these energies.

Note that $4 \sigma_{\rm R}(p+{\rm C})$ is appreciably larger than
$\sigma_{\rm R}$ for $^{12}$C+$^{12}$C reactions 
for incident energies per nucleon less than about 100 MeV. 
This implies that the classical picture, 
the geometrical description of the cross section
underlying Eq.\ (\ref{sigbs2}), breaks down. 
If we could adopt $4 \sigma_{\rm R}(p+{\rm C})$ as a basis 
in assessing $\sigma_{\rm R}$ for 
$^{12}$C+$^{12}$C reactions, we need a certain 
``transparency" effect \cite{Kox} to reproduce the data.
We remark that the Coulomb effect on $\sigma_{\rm R}$
would be hard to resolve the disagreement alone, 
because the cross-section reduction 
due to the Coulomb repulsion between the projectile and target
nuclei is stronger 
for proton-carbon reactions than for carbon-carbon ones
at given incident energy per nucleon. 
Similar deviations also appear 
for $^9$Be+$^{12}$C and $^{27}$Al+$^{12}$C cases 
of incident energy per nucleon lower than about 100 MeV
as in Fig.~\ref{bealonc}.

We should be careful about a deviation of $\sigma_{\rm BS}$ 
from 4$\sigma_{\rm R}(p+{\rm C})$ that is appreciable around
500 MeV per nucleon in Fig.~\ref{ccsigr}. It does not imply a
flaw in the black-sphere model, but simply reflects
the fact that the measured values of $\sigma_{\rm R}(p+{\rm C})$ at
proton incident energies of 220--570 MeV \cite{ren:npa}
are larger than expected from the systematics \cite{IKO}.  
Note that a similar tendency appears in the values of
$\sigma_{\rm BS}$ for $^{9}$Be+$^{12}$C and 
$^{27}$Al+$^{12}$C reactions 
derived from the same values of $\sigma_{\rm R}(p+{\rm C})$
(see Fig.~\ref{bealonc}).


     In summary, we have pointed out that, 
for stable nuclei incident on a carbon target, 
there is a significant difference between 
real $\sigma_{\rm I}$ data 
and pseudo $\sigma_{\rm R}$ data even at relativistic energies, 
especially at small mass number. 
This difference would lead to possible uncertainties
of about 0--0.3 fm in estimates of nuclear matter radii,
if relying on the $\sigma_{\rm I}$ data alone.
We have found that 
this difference is consistent with the fact that 
$\sigma_{\rm I} < \sigma_{\rm R}$ 
and generally stays within $0-100$ mb.
The difference is clear for small $A$ while 
it is less clear for larger $A$.
This implies that the scale $\Delta a$ characterizing
the difference between the black-sphere and interaction radii
of the projectile nucleus decreases 
no slower than $A^{-1/3}$ as $A$ increases, 
a feature relevant to the problem of what energy
scale controls the breakup of the projectile.
Of course, the above implications strongly depend on 
the validity of our black-sphere picture, 
which is based on $\sigma_{\rm R} \cong \sigma_{\rm BS}$. 
The presently available $\sigma_{\rm R}$ data
for $^{9}$Be, $^{12}$C, and $^{27}$Al incident on $^{12}$C 
support the above relation.

We acknowledge T. Motobayashi for his 
constructive comments and encouragement
during the course of this work, 
K. Yabana for his critical comments, 
and also M. Takechi and M. Fukuda for 
providing us with the latest data 
for the total reaction cross sections 
of $^{9}$Be, $^{12}$C, and $^{27}$Al incident on $^{12}$C. 
We also acknowledge the members of Japan Charged-Particle 
Nuclear Reaction Data Group (JCPRG), especially N. Otuka, 
for kindly helping us collect various data sets. 
A.K. would like to thank H. Sakurai for stimulating this work.

\end{document}